\documentclass[a4paper]{article}
\usepackage{graphicx}
\usepackage{amsmath}
\usepackage{amssymb}
\usepackage[sort&compress,numbers]{natbib}
\usepackage{url}
\usepackage[a4paper, margin=1in]{geometry}

\newcommand{\lp}[1]{\left( #1 \right)}
\newcommand{\lb}[1]{\left[ #1 \right]}
\newcommand{\lc}[1]{\left\{ #1 \right\}}

\begin{document}

\title{Costly punishment sustains indirect reciprocity under low defection detectability}

\author{
  Yohsuke Murase\thanks{RIKEN Center for Computational Science, Kobe, 650-0047, Japan, \texttt{yohsuke.murase@gmail.com}}
}
\maketitle

\begin{abstract}
  Cooperation is fundamental to human societies, and indirect reciprocity, where individuals cooperate to build a positive reputation for future benefits, plays a key role in promoting it.
  Previous theoretical and experimental studies have explored both the effectiveness and limitations of costly punishment in sustaining cooperation.
  While empirical observations show that costly punishment by third parties is common, some theoretical models suggest it may not be effective in the context of indirect reciprocity, raising doubts about its potential to enhance cooperation.
  In this study, we theoretically investigate the conditions under which costly punishment is effective.
  Building on a previous model, we introduce a new type of error in perceiving actions, where defection may be mistakenly perceived as cooperation.
  This extension models a realistic scenario where defectors have a strong incentive to disguise their defection as cooperation.
  Our analysis reveals that when defection is difficult to detect, norms involving costly punishment can emerge as the most efficient evolutionarily stable strategies.
  These findings demonstrate that costly punishment can play a crucial role in promoting cooperation within indirect reciprocity.
\end{abstract}

\section{Introduction}

Cooperation is a fundamental aspect of human societies, enabling individuals to work together and achieve outcomes that would be difficult or impossible to accomplish alone~\cite{rand:TCS:2013,melis:ptrs:2010}.
One of the key mechanisms that sustains cooperation is indirect reciprocity, where individuals gain a reputation based on their behavior, influencing their future interactions with others~\cite{nowak2005evolution,wedekind2002long,sigmund2012moral,santos2021complexity,okada2020review}.
In this framework, those who cooperate tend to receive a good reputation, which encourages others to cooperate with them in subsequent interactions.
Conversely, individuals who defect acquire a bad reputation, which can lead to a long-term loss in future interactions.

Punishment is often used as a strategy to deter defection and promote cooperation~\cite{fehr2002altruistic,de2004neural,mathew2011punishment,wu2022reward,balafoutas2014direct,ule2009indirect}.
By imposing penalties on defectors, groups can maintain or even enhance cooperation.
Numerous behavioral experiments and real-world observations show that costly punishment---where individuals incur personal costs to penalize defectors---is a common phenomenon in human societies.
People often punish perceived wrongdoers even at a cost to themselves, reflecting a deep-rooted willingness to uphold social norms.

However, some theoretical studies question the effectiveness of punishment in promoting cooperation~\cite{guala2012reciprocity,wu2022reward,raihani2015reputation,raihani2019punishment,li2018punishment,hauser2014punishment,rand2011evolution}.
Ohtsuki et al.~\cite{ohtsuki2009indirect} have suggested that the effectiveness of costly punishment may be limited in the context of indirect reciprocity.
Their study indicates that the introduction of costly punishment often results in lower overall payoffs compared to scenarios where punishment is not used.
Moreover, in environments that are too noisy or where the cost of punishment is excessively high, costly punishment can become less efficient than strategies like always defecting (ALLD).
As a result, the conditions under which costly punishment is effective appear to be quite narrow, leading to the conclusion that cooperation and defection alone may suffice to maintain cooperative behavior.
However, their model assumes that players' actions are always perfectly observed, which is often unrealistic in real-world scenarios.
This raises the critical question of whether the detectability of defection itself may influence the effectiveness of costly punishment in indirect reciprocity.

Our study addresses this question by exploring how costly punishment fosters cooperation when defection is difficult to detect.
In reality, defectors may attempt to disguise their actions as cooperative, while cooperators strive to ensure their behavior is recognized to avoid unfair punishment.
This creates an asymmetry: while cooperation is often observable, defection may go unnoticed or be misinterpreted.

To incorporate this asymmetry in imperfect monitoring, we extend the model of Ohtsuki et al.~\cite{ohtsuki2009indirect} by introducing the concept of the ``detectability of defection'', a parameter that quantifies the difficulty in identifying defection accurately.
Our theoretical analysis reveals that social norms incorporating costly punishment emerge as the most efficient evolutionarily stable strategies under a broad range of conditions, particularly when defection is hard to detect.
When defection detectability is low, maintaining cooperation becomes more challenging, but costly punishment helps sustain cooperation by countering undetected defection.

These findings contrast with previous studies that focused on errors in distinguishing good and bad reputations~\cite{ohtsuki2009indirect,nakamura2011indirect,ohtsuki2004should,ohtsuki2006leading,murase2023indirect}.
In our model, the challenge lies in distinguishing defection from cooperation, leading to a starkly different outcome.
This highlights the importance of considering the nature of errors when evaluating the role of costly punishment in indirect reciprocity.

\section{Model}

\begin{figure}
  \centering
  \includegraphics[width=0.9\textwidth]{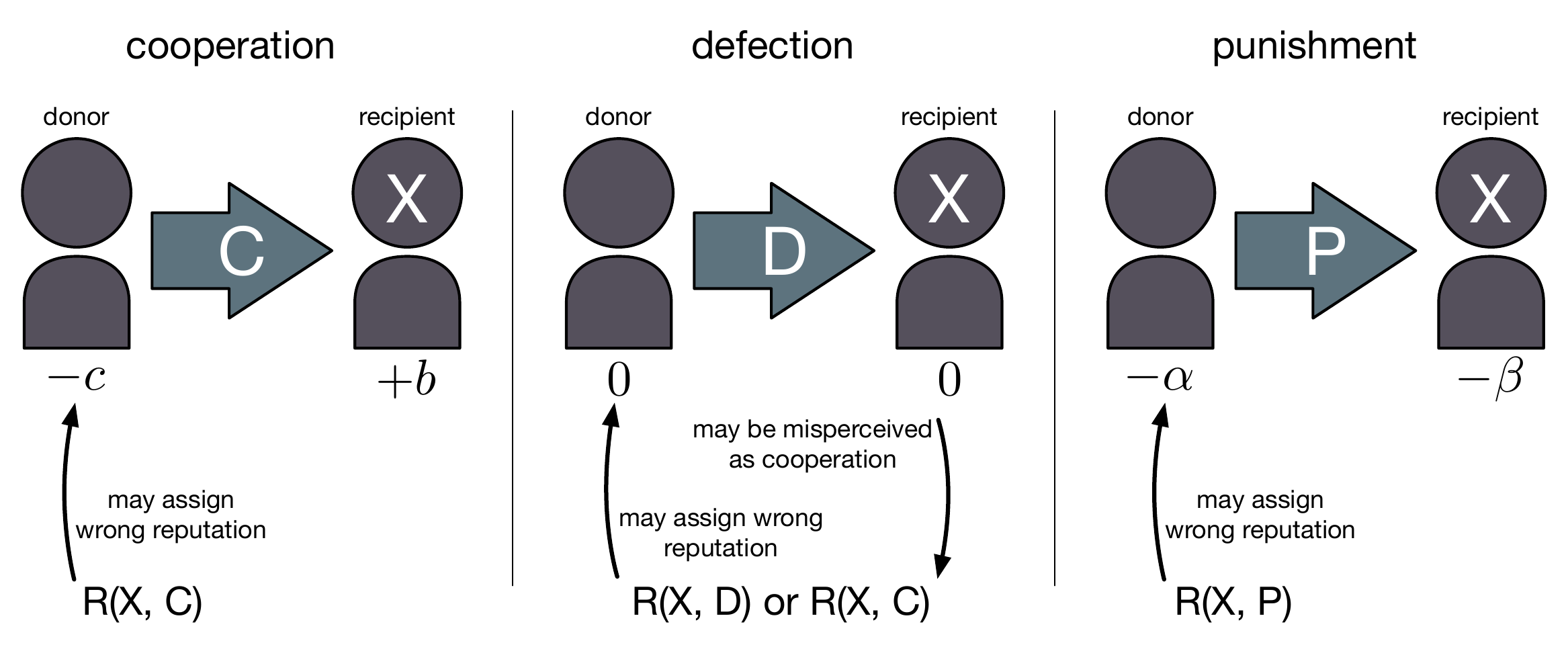}
  \caption{
    A schematic diagram of the model.
    At each step, a donor and a recipient are randomly selected.
    The donor then chooses an action from three options: cooperation (C), defection (D), or punishment (P).
    If the donor chooses to cooperate, the recipient receives a benefit $b$, while the donor incurs a cost $c$.
    In the case of defection, both players receive a payoff of zero.
    If the donor opts for punishment, the recipient incurs a loss $\beta$, and the donor pays a punishment cost $\alpha$.
    The donor's action influences their reputation as perceived by the recipient.
    The new reputation depends on the recipient's current reputation $X$ and the donor's action.
    If the donor defects, there is a probability $\epsilon_{DC}$ that the action will be misperceived as cooperation.
    In all other cases, the action is correctly identified.
    Additionally, the assessment process is prone to errors.
    With a probability $\mu$, the opposite reputation is assigned to the donor.
  }
  \label{fig:schematic}
\end{figure}

\begin{table}
  \centering
  \begin{tabular}{|c|cc|cccc|}
  \hline
        & \multicolumn{2}{c|}{Action Rule} & \multicolumn{4}{c|}{Assessment Rule}     \\
  Name  & $S(G)$ & $S(B)$                   & $R(G,C)$ & $R(G,D)$ & $R(B,C)$ & $R(B,D)$ \\
  \hline
  Image Scoring   & $C$ & $D$ & $1$ & $0$ & $1$ & $0$ \\
  Simple Standing & $C$ & $D$ & $1$ & $0$ & $1$ & $1$ \\
  Stern Juddging  & $C$ & $D$ & $1$ & $0$ & $0$ & $1$ \\
  ALLC            & $C$ & $C$ & $*$ & $*$ & $*$ & $*$ \\
  ALLD            & $D$ & $D$ & $*$ & $*$ & $*$ & $*$ \\
  \hline
  \end{tabular}
  \caption{
    Well-known examples of social norms without costly punishment.
    The action rule prescribes which action to take when the recipient is assessed as good ($G$) or bad ($B$).
    The assessment rule $R(X, A)$ prescribes the probability of assigning a good reputation to the donor, depending on the recipient's reputation $X$ and the donor's action $A$.
    An asterisk ($*$) indicates that the value can be arbitrary.
    Simple Standing and Stern Judging norms are the only ESS norms that achieve full cooperation in the low-error limit $\mu \to 1$.
    These are two of the ``leading eight'' norms.
  }
  \label{table_example}
\end{table}

In this study, we follow the basic framework established by Ohtsuki et al.~\cite{ohtsuki2009indirect}.
We consider an infinitely large population of players, where individuals repeatedly interact in pairwise donation games.
In each round, two players are randomly chosen, one as the donor and the other as the recipient.
The donor chooses an action from cooperation (C), defection (D), or punishment (P).
Cooperation incurs a cost $c > 0$ to the donor and provides a benefit $b > c$ to the recipient.
Defection results in a payoff of zero for both players.
Punishment incurs a cost $\alpha > 0$ to the donor and imposes a loss of $\beta > 0$ on the recipient.
If the donation game is played only once, the donor benefits more by defecting, creating a social dilemma.
However, we consider the scenario where members of the population repeatedly engage in many donation games with changing donor-recipient pairs.
During these interactions, players build reputations based on their past actions, which may influence future cooperation.
Following the standard convention in public assessment models~\cite{ohtsuki2004should,ohtsuki2006leading,murase2023indirect}, we assume that reputations are binary, with players being either good ($G$) or bad ($B$), and that each player's reputation is publicly shared without disagreement.

How players' reputations are formed and how they act based on these reputations depend on their social norms.
In our study, a social norm consists of an action rule $S$ and an assessment rule $R$, as illustrated in Fig.~\ref{fig:schematic}.
The action rule $S(X)$ determines the action---either $C$, $D$, or $P$---when acting as a donor.
This choice may depend on the recipient's reputation $X \in \{G, B\}$.
The action rule is defined as a map $S: X \rightarrow A$, where $X \in \{G, B\}$ represents the recipient's reputation and $A \in \{C, D, P\}$ represents the donor's action.
Additionally, we consider an assessment rule $R(X, A)$, which determines the probability of assigning a good reputation to the recipient based on the donor's action and the recipient's reputation.
In this study, we consider only deterministic norms, so $R$ is constrained to be either $0$ or $1$; that is, $R: X \times A \rightarrow \{0, 1\}$.
We consider the so-called "second-order" social norms, where the assessment rules depend on both the recipient's reputation and the donor's action.
As a reference, we list several well-known norms without costly punishment in Table~\ref{table_example}.
Hereafter, we denote a norm with $S(G) = C$ and $S(B) = D$ as a CD norm, a norm with $S(G) = C$ and $S(B) = P$ as a CP norm, and so on.

Reputation dynamics are subject to errors.
In our model, we consider two types of errors: assessment errors and detection errors.
An assessment error, also conventionally introduced in previous studies~\cite{ohtsuki2004should,ohtsuki2006leading,ohtsuki2009indirect,murase2023indirect}, may occur when a new reputation is assigned to the donor.
With probability $\mu \ (\leq 1/2)$, the opposite reputation is assigned to the donor.
This can equivalently be interpreted as the population's ability to distinguish between good and bad reputations with probability $q = 1 - 2\mu$~\cite{ohtsuki2009indirect}.
The parameter $q$ quantifies the ability to distinguish between good and bad reputations and is called the 'social resolution'.

The second type of error is detection error, which is newly introduced in this study.
Actions are not always correctly observed and may be misperceived.
Here, we consider the case where defection is perceived as cooperation, due to the natural incentive for defectors to disguise their defection as cooperation.
When a donor defects, the action is perceived as cooperation with probability $\epsilon_{DC}$ and correctly identified as defection with probability $1 - \epsilon_{DC}$.
We define the detectability of defection as $\xi = 1 - \epsilon_{DC}$.
When $\epsilon_{DC} = 0$ ($\xi = 1$), defections are perfectly detected, corresponding to the baseline model~\cite{ohtsuki2009indirect}.
When $\epsilon_{DC} = 1$ ($\xi = 0$), defections are indistinguishable from cooperation and are always misperceived as cooperation.

Given these two types of errors, the effective assessment rule $\tilde{R}(X, A)$~\cite{murase2023indirect} is expressed as follows:
\begin{equation}
\begin{cases}
  \tilde{R}(X, C) &= (1 - 2\mu) R(X, C) + \mu   \\
  \tilde{R}(X, D) &= (1 - 2\mu) [ \epsilon_{DC} R(X, C) + (1 - \epsilon_{DC}) R(X, D) ] + \mu \\
  \tilde{R}(X, P) &= (1 - 2\mu) R(X, P) + \mu
\end{cases}.
\label{eq:effective_assessment_rule}
\end{equation}
For any $X$ and $A$, $\mu \leq \tilde{R}(X, A) \leq 1 - \mu$, ensuring that the average reputation converges to a unique equilibrium value regardless of the initial state when $\mu > 0$~\cite{ohtsuki2004should,ohtsuki2006leading,murase2023indirect}.

All subsequent analysis reproduces the findings of the baseline model~\cite{ohtsuki2009indirect} when $\xi = 1$.

\section{Analysis}

In the following, we derive the norms that are evolutionarily stable strategies (ESS) within the model described above.
Among these ESS norms, we identify the most efficient ones---those leading to the highest average payoff at equilibrium---for the given environmental parameters $\{b, c, \alpha, \beta, q, \xi\}$.

ESS norms are defined as those that satisfy the following condition: A single mutant player (one with a different action rule $S'$) has a strictly lower payoff than the players following the resident action rule $S$ for all $S' \neq S$.
Obviously, any DD norms (ALLD), are always ESS.
We only need to consider CD and CP norms, as these are the only norms that can be ESS and more efficient than DD norms.
This is because CC (ALLC) and PP (ALLP) norms are never ESS, and DP norms are always less efficient than DD norms.
We do not need to consider the other norms (DC, PC, and PD) since they are equivalent to the above norms by simply swapping the labels of $G$ and $B$.

\subsection{ESS condition for CD norms}

First, we derive the condition for CD norms to be ESS.
In the previous work~\cite{ohtsuki2009indirect}, the Bellman equation was derived to identify ESS norms.
Here, we use a simpler approach to identify ESS norms, inspired by Ref.~\cite{murase2023indirect}.

Suppose the population follows a CD norm.
We consider the best response in each context, meaning the action that maximizes the donor's payoff.
Here, the context refers to the recipient's reputation.
For instance, we compare the expected payoffs when the donor chooses $C$, $D$, or $P$ against a $G$-recipient.
If the expected payoff from $C$ is higher than that from $D$ or $P$, the donor's best response against a $G$-recipient is $C$.
The same comparison is made for $B$-recipients.
If the best response against a $B$-recipient is $D$, then the resident norm is the best response to itself, making the norm ESS.

To compare the expected payoff for each action, we first calculate the value of the $G$ reputation.
Consider a player currently assessed as $G$.
In subsequent interactions, the player will be chosen as a donor with probability $1/2$ and as a recipient with the same probability.
This player receives a benefit $b$ as a recipient as long as they are assessed as $G$ and typically has one interaction as a recipient before being chosen as a donor.
When they are chosen as a donor, a new reputation is assigned, which is completely independent of the current reputation since the social norm is second-order.
Thus, being assessed as $G$ is worth $b$ more than being assessed as $B$.

Now, comparing actions is straightforward.
First, we compare $C$ and $D$ against $G$-recipients.
In the subsequent interaction as a recipient, a cooperating donor receives a benefit $b$ with probability $\tilde{R}\lp{G, C}$ and a payoff of zero with probability $1 - \tilde{R}\lp{G, C}$.
The total expected payoff over two rounds (one as a donor and one as a recipient) is $\tilde{R}\lp{G, C} b - c$.
In contrast, the expected payoff of defection is similarly expressed as $\tilde{R}\lp{G, D} b$.
Thus, cooperation is better than defection if and only if $\tilde{R}\lp{G, C} b - c > \tilde{R}\lp{G, D} b$.
Similarly, we compare $C$ and $P$ against $G$-recipients, $D$ and $C$ against $B$-recipients, and $D$ and $P$ against $B$-recipients.
The ESS conditions for CD norms are summarized by the following four inequalities:
\begin{align}
  \lb{ \tilde{R}\lp{G, C} - \tilde{R}\lp{G, D} }b &> c          \\
  \lb{ \tilde{R}\lp{G, C} - \tilde{R}\lp{G, P} }b &> c - \alpha \\
  \lb{ \tilde{R}\lp{B, D} - \tilde{R}\lp{B, C} }b &> -c         \\
  \lb{ \tilde{R}\lp{B, D} - \tilde{R}\lp{B, P} }b &> -\alpha
\end{align}
These are simplified as the following, respectively:
\begin{align}
  & q \xi \lb{R\lp{G, C} - R\lp{G, D} } b > c               \label{eq:cd_norm_gc_gd} \\
  & q     \lb{R\lp{G, C} - R\lp{G, P} } b > c - \alpha      \label{eq:cd_norm_gc_gp} \\
  & q \xi \lb{R\lp{B, D} - R\lp{B, C} } b > -c              \label{eq:cd_norm_bd_bc} \\
  & q \xi \lb{R\lp{B, D} - R\lp{B, C} } b + q \lb{R\lp{B, C} - R\lp{B, P} } b > -\alpha    \label{eq:cd_norm_bd_bp}
\end{align}

From Eqs.~(\ref{eq:cd_norm_gc_gd}) and (\ref{eq:cd_norm_gc_gp}), the following conditions are necessary for a norm to be ESS:
\begin{align}
  R(G, C) &= 1                              \label{eq:cd_norm_ess_condition1}  \\
  R(G, D) &= 0                              \label{eq:cd_norm_ess_condition2}  \\
  R(G, P) &= \begin{cases}
    0 & \text{if } c \geq \alpha \\
    \text{any} & \text{if } c < \alpha
    \end{cases}                             \label{eq:cd_norm_ess_condition3} \\
  q \xi b &> c                              \label{eq:cd_norm_ess_condition4}
\end{align}
Norms that satisfy the remaining conditions in Eqs.~(\ref{eq:cd_norm_bd_bc}), (\ref{eq:cd_norm_bd_bp}), and (\ref{eq:cd_norm_ess_condition4}) are listed in Table~\ref{table:cd_norm_ess_conditions}.

\begin{table}[h]
  \centering
  \begin{tabular}{|ccc|c|c|}
  \hline
  $R(B, C)$ & $R(B, D)$ & $R(B, P)$ & Condition & Cooperation level \\ \hline
  1         & 1         & any       & any                      & $1 - \mu$  \\
  0         & 1         & 1         & $q (1 - \xi) b < \alpha$ & $1 - \mu / \lb{1 - \lp{1 - 2\mu}\epsilon_{DC} }$ \\
  0         & 1         & 0         & any                      & $1 - \mu / \lb{1 - \lp{1 - 2\mu}\epsilon_{DC} }$  \\
  0         & 0         & 1         & $qb < \alpha$            & $1/2$  \\
  0         & 0         & 0         & any                      & $1/2$  \\ \hline
  \end{tabular}
  \caption{
    Social norms that satisfy the conditions in Eqs.~(\ref{eq:cd_norm_bd_bc}), (\ref{eq:cd_norm_bd_bp}), and (\ref{eq:cd_norm_ess_condition4}) are listed.
    The conditions on the environmental parameters $q$, $\xi$, $b$, $c$, and $\alpha$, which are additional requirements to the condition in Eq.~(\ref{eq:cd_norm_ess_condition4}), are shown in the middle column.
    The cooperation level calculated from Eq.~(\ref{eq:equilibrium_reputation_cd}) is displayed in the rightmost column.
  }
  \label{table:cd_norm_ess_conditions}
\end{table}

The cooperation level for these ESS norms is calculated.
The average reputation (the fraction of $G$-players), denoted as $h$, follows this time evolution equation:
\begin{equation}
  \dot{h} = h \tilde{R}\lp{G, S\lp{G} } + \lp{1 - h} \tilde{R}\lp{B, S\lp{B}} - h.
  \label{eq:reputation_evolution}
\end{equation}
Thus, the equilibrium reputation $h^{\ast}$ for CD norms, which corresponds to the cooperation level, is given by
\begin{equation}
  h^{\ast} = \frac{ \tilde{R}\lp{B, D} }{ 1 - \tilde{R}\lp{G, C} + \tilde{R}\lp{B, D} }.
  \label{eq:equilibrium_reputation_cd}
\end{equation}
Since $\tilde{R}(G,C) = 1 - \mu$ from Eq.~(\ref{eq:cd_norm_ess_condition1}), the cooperation levels for the norms in Table~\ref{table:cd_norm_ess_conditions} are calculated and presented in the rightmost column.

Therefore, the norm at the top of Table~\ref{table:cd_norm_ess_conditions} has the highest cooperation level and the widest parameter region.
This norm, which is similar to Simple Standing, is the most efficient because the entry $R(B, C) = 1$ does not hurt the reputation under detection error.

In summary, ESS CD norms achieve the highest cooperation level of $1 - \mu$ when $q \xi b > c$.
Otherwise (i.e., when $q \xi b \leq c$), no CD norms are ESS.

\subsection{ESS condition for CP norms}

Next, we consider CP norms.
Similar to the CD norms, we compare the actions in each context.
The ESS conditions are expressed as the following four inequalities:
\begin{align}
  \lb{ \tilde{R}\lp{G, C} - \tilde{R}\lp{G, D} } \lp{b+\beta} &> c \\
  \lb{ \tilde{R}\lp{G, C} - \tilde{R}\lp{G, P} } \lp{b+\beta} &> c - \alpha \\
  \lb{ \tilde{R}\lp{B, P} - \tilde{R}\lp{B, C} } \lp{b+\beta} &> \alpha - c \\
  \lb{ \tilde{R}\lp{B, P} - \tilde{R}\lp{B, D} } \lp{b+\beta} &> \alpha
\end{align}
Note that under CP norms, being assessed as $G$ is worth $\lp{b + \beta}$ more than being assessed as $B$, since a $B$ recipient is punished by $-\beta$.
These inequalities can be simplified as follows:
\begin{align}
  & q \xi \lb{ R\lp{G, C} - R\lp{G, D} } \lp{b+\beta} > c              \label{eq:cp_norm_gc_gd} \\
  & q     \lb{ R\lp{G, C} - R\lp{G, P} } \lp{b+\beta} > c - \alpha     \label{eq:cp_norm_gc_gp} \\
  & q     \lb{ R\lp{B, P} - R\lp{B, C} } \lp{b+\beta} > \alpha - c     \label{eq:cp_norm_bp_bc} \\
  & q \lc{ \lb{ R\lp{B, P} - R\lp{B, C} } + \xi \lb{ R\lp{B, C} - R\lp{B, D} } } \lp{b+\beta} > \alpha  \label{eq:cp_norm_bp_bd}
\end{align}

From Eqs.~(\ref{eq:cp_norm_gc_gd}) and (\ref{eq:cp_norm_gc_gp}), the following conditions are necessary for a norm to be ESS:
\begin{align}
  R(G, C) &= 1                               \label{eq:cp_norm_ess_condition1} \\
  R(G, D) &= 0                               \label{eq:cp_norm_ess_condition2} \\
  R(G, P) &= \begin{cases}
    0 & \text{if } c \geq \alpha \\
    \text{any} & \text{if } c < \alpha
    \end{cases}                              \label{eq:cp_norm_ess_condition3} \\
  q \xi \lp{b + \beta} &> c                  \label{eq:cp_norm_ess_condition4}
\end{align}
Norms that satisfy the remaining conditions from Eqs.~(\ref{eq:cp_norm_bp_bc}) and (\ref{eq:cp_norm_bp_bd}) are listed in Table~\ref{table:cp_norm_ess_conditions}.

\begin{table}[h]
  \centering
  \begin{tabular}{|ccc|c|c|}
  \hline
  $R(B, C)$ & $R(B, D)$ & $R(B, P)$ & Condition & Cooperation level \\ \hline
  0         & 0         & 1         & $q (b + \beta) > \alpha$                          & $1 - \mu$ \\
  0         & 1         & 1         & $q (1 - \xi) (b + \beta) > \alpha$                & $1 - \mu$ \\
  1         & 0         & 1         & $c > \alpha$ and $q \xi (b+\beta) > \alpha$       & $1 - \mu$ \\
  \hline
  \end{tabular}
  \caption{
    Social norms that satisfy the conditions in Eqs.~(\ref{eq:cp_norm_bp_bc}) and (\ref{eq:cp_norm_bp_bd}) are listed.
    The conditions on the environmental parameters $q$, $\xi$, $b$, $c$, and $\alpha$, which are additional requirements to the condition in Eq.~(\ref{eq:cd_norm_ess_condition4}), are provided.
    The cooperation level calculated from Eq.~(\ref{eq:equilibrium_reputation_cp}) is displayed in the rightmost column.
  }
  \label{table:cp_norm_ess_conditions}
\end{table}

The equilibrium reputation $h^{\ast}$, which is equal to the cooperation level, is obtained from Eq.~(\ref{eq:reputation_evolution}) as follows:
\begin{equation}
  h^{\ast} = \frac{ \tilde{R}\lp{B, P} }{ 1 - \tilde{R}\lp{G, C} + \tilde{R}\lp{B, P} }.
  \label{eq:equilibrium_reputation_cp}
\end{equation}
Since $\tilde{R}(G,C) = 1 - \mu$ from Eq.~(\ref{eq:cp_norm_ess_condition1}) and $\tilde{R}(B,P) = 1 - \mu$ from Table~\ref{table:cp_norm_ess_conditions}, the cooperation levels for these norms are $1 - \mu$.

The norm at the top of Table~\ref{table:cp_norm_ess_conditions} has the widest parameter region.
This norm is similar to the Stern Judging norm in that any action not prescribed by the norm always leads to a bad reputation.

In summary, ESS norms exist when $q \xi (b + \beta) > c$ and $q(b + \beta) > \alpha$.
Otherwise, CP norms cannot be ESS.

\subsection{The most efficient ESS norms}

The equilibrium payoffs under the most efficient ESS norms obtained above are given by:
\begin{align}
  \pi_{CD} &= (1 - \mu) (b - c) \\
  \pi_{CP} &= (1 - \mu) (b - c) - \mu (\alpha + \beta) \\
  \pi_{DD} &= 0
\end{align}
Since $\pi_{CD} > \pi_{CP}$ and $\pi_{CD} > \pi_{DD}$, CD norms are the most efficient if they exist.
The comparison between CP norms and DD norms is more intricate.
CP norms are more efficient than DD norms ($\pi_{CP} > \pi_{DD}$) when:
\begin{equation}
  q > \frac{-b + c + \alpha + \beta}{b - c + \alpha + \beta}
  \label{eq:cp_more_efficient_than_dd}
\end{equation}

Therefore, the most efficient ESS norms are
\begin{equation}
  \begin{cases}
  \text{CD norms} & \text{when } q \xi b > c  \\
  \text{CP norms} & \text{when } q \xi b \leq c \text{ and } q \xi (b + \beta) > c \text{ and } q(b + \beta) > \alpha \text{ and } q > \frac{-b + c + \alpha + \beta}{b - c + \alpha + \beta}  \\
  \text{DD norms} & \text{otherwise}
  \end{cases}
  \label{eq:most_efficient_norms}
\end{equation}

\begin{figure}
  \centering
  \includegraphics[width=0.9\textwidth]{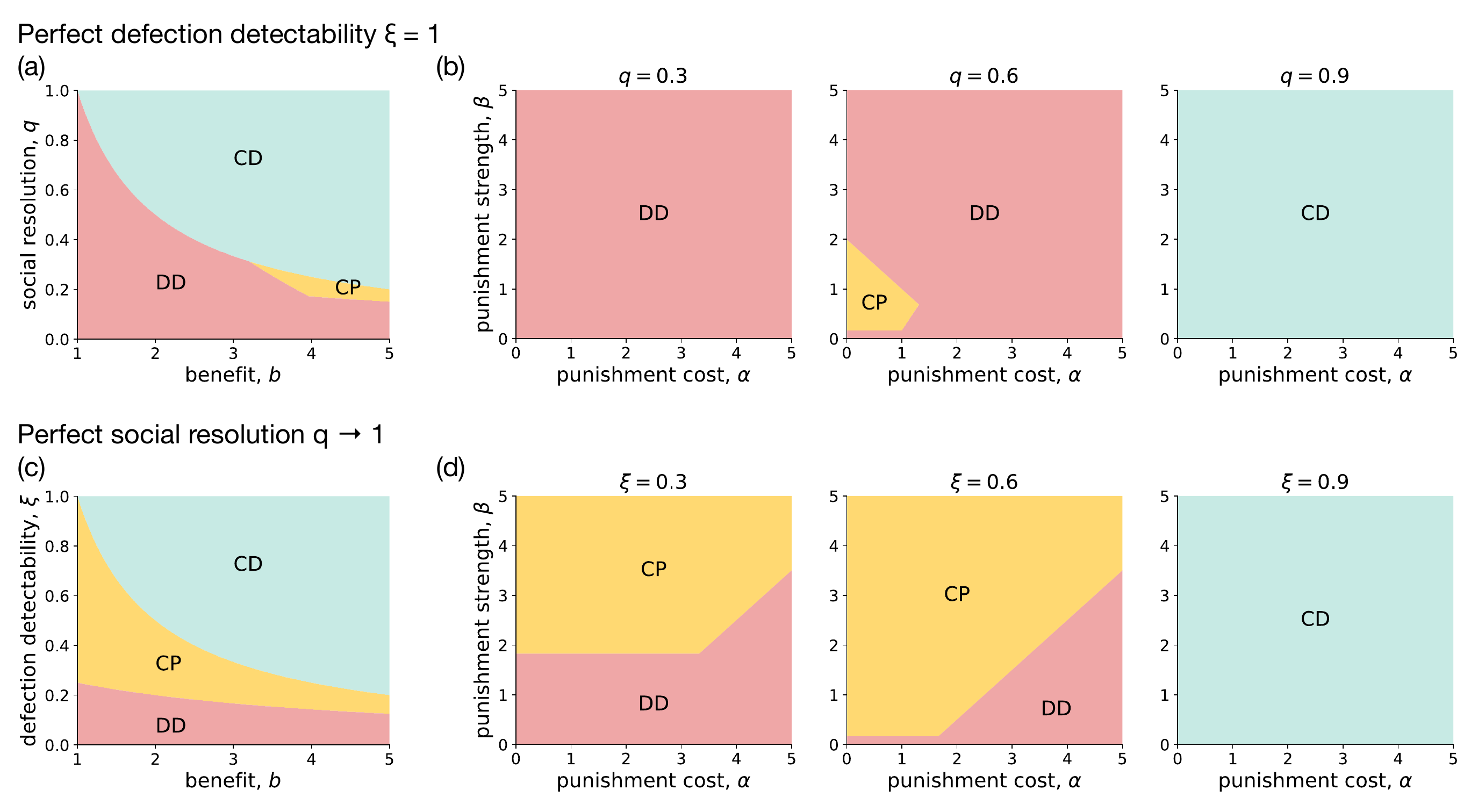}
  \caption{
    The parameter regions where the most efficient ESS norms are CD norms, CP norms, and DD norms.
    The upper panels (a, b) show cases with perfect defection detectability ($\xi = 1$), while the lower panels (c, d) depict cases with perfect social resolution ($q \to 1$).
    The regions are derived from Eq.~(\ref{eq:most_efficient_norms_xi1}) for the upper panels and Eq.~(\ref{eq:most_efficient_norms_q1}) for the lower panels.
    The left panels (a) and (c) display projections in the $b$-$q$ and $b$-$\xi$ planes, respectively, for $\alpha = 1.2$ and $\beta = 3$.
    The right panels (b) and (d) display projections in the $\alpha$-$\beta$ planes for different values of $q$ and $\xi$, respectively.
  }
  \label{fig:CD_CP_DD_efficiency}
\end{figure}
With perfect defection detectability ($\xi = 1$), the model reproduces the baseline model~\cite{ohtsuki2009indirect}.
As derived in previous work, the most efficient norms are
\begin{equation}
  \begin{cases}
  \text{CD norms} & \text{when } q b > c  \\
  \text{CP norms} & \text{when } q b \leq c \text{ and } q (b + \beta) > c \text{ and } q(b + \beta) > \alpha \text{ and } q > \frac{-b + c + \alpha + \beta}{b - c + \alpha + \beta}  \\
  \text{DD norms} & \text{otherwise}
  \end{cases}
  \label{eq:most_efficient_norms_xi1}
\end{equation}
These regions are depicted in the top panels of Fig.~\ref{fig:CD_CP_DD_efficiency}.
The regions where CP norms are the most efficient are small, and in most cases, the most efficient ESS norms are either CD norms or DD norms.

On the other hand, with perfect social resolution ($q \to 1$), the most efficient ESS norms are:
\begin{equation}
  \begin{cases}
  \text{CD norms} & \text{when } \xi b > c  \\
  \text{CP norms} & \text{when } \xi b \leq c \text{ and } \xi (b + \beta) > c \text{ and } (b + \beta) > \alpha  \\
  \text{DD norms} & \text{otherwise}
  \end{cases}
  \label{eq:most_efficient_norms_q1}
\end{equation}
In this case, the parameter region where CP norms are the most efficient is much larger than in the baseline model, as depicted in the lower panels of Fig.~\ref{fig:CD_CP_DD_efficiency}.
Furthermore, CP norms become as efficient as CD norms when $q \to 1$.

\begin{figure}
  \centering
  \includegraphics[width=0.9\textwidth]{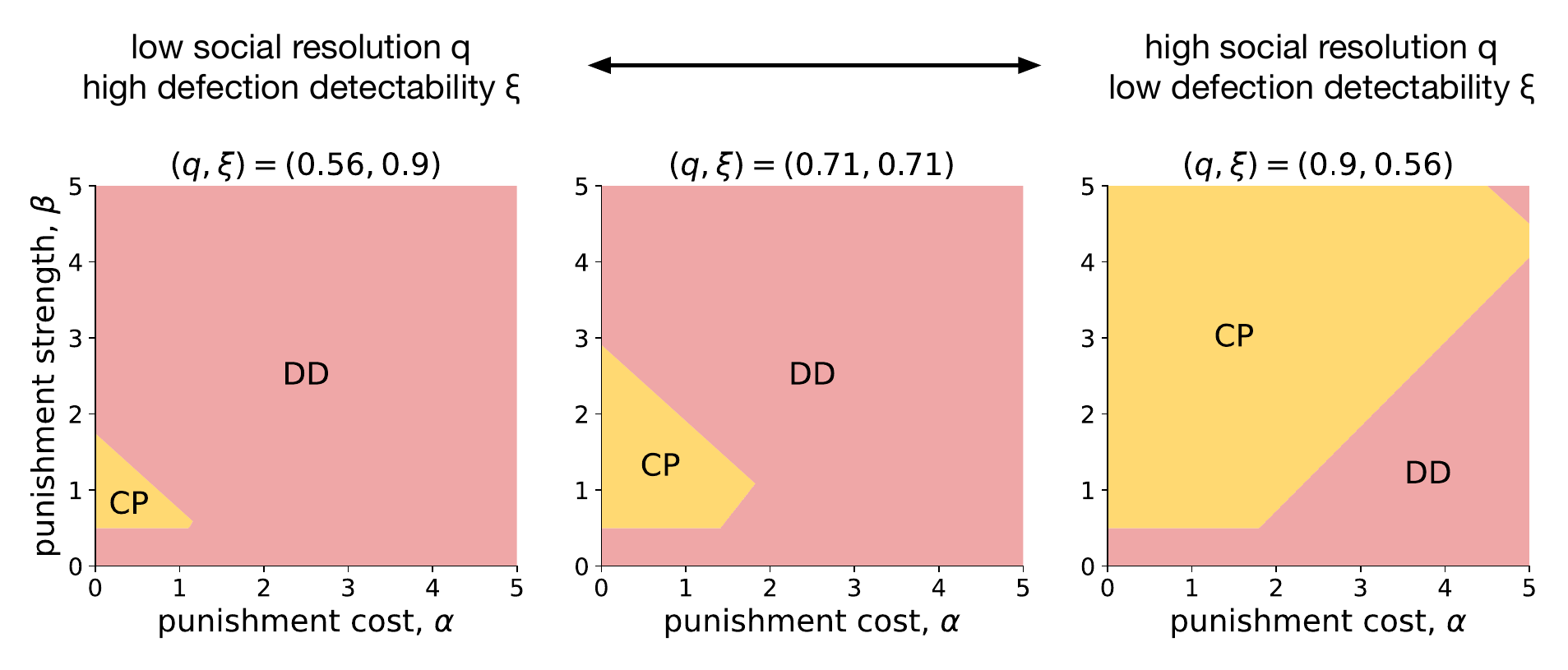}
  \caption{
    The same figure as Fig.~\ref{fig:CD_CP_DD_efficiency}(b) and (d), but with intermediate values of $\xi$ and $q$, while keeping $q \xi = 0.5$ and $b = 1.5$ fixed.
From left to right, the values are $(q, \xi) = (5/9, 9/10)$, $(1/\sqrt{2}, 1/\sqrt{2})$, and $(9/10, 5/9)$.
These regions are calculated using Eq.~(\ref{eq:most_efficient_norms}).
  }
  \label{fig:sweep_xi_q}
\end{figure}

A crossover between these two limiting cases is shown in Fig.~\ref{fig:sweep_xi_q} to demonstrate that these behaviors are not unique to the limiting cases ($\xi = 1$ or $q \to 1$).
In this figure, the most efficient norms are shown for intermediate values of $\xi$ and $q$ while keeping the overall error rate fixed at $q \xi = 0.5$.
As shown in this figure, the region where CP norms are effective gradually expands as $\xi$ decreases and $q$ increases.
Furthermore, as $q$ increases, the cooperation level in CP norms improves and approaches that in CD norms.
Thus, high social resolution is crucial for the effectiveness of CP norms.

The stark difference between high $\xi$ and high $q$ can be understood as follows.
First, when both $\xi$ and $q$ are high, CD norms are sufficient to maintain evolutionarily stable cooperation through Simple Standing or Stern Judging norms.
Punishment may be effective when the environment is noisy, either due to low $\xi$ or low $q$.
In these cases, the population cannot accurately identify defectors, making punishment necessary to sufficiently penalize defectors and stabilize cooperation.
However, when $q$ is low, the cooperation level is significantly reduced because the population includes individuals with bad reputations.
When players punish bad recipients under CP norms, the equilibrium payoffs are significantly reduced, eventually becoming less efficient than even DD norms.
In summary, while costly punishment stabilizes cooperation in noisy environments, it sacrifices the efficiency of cooperation.
As a result, CP norms rarely emerge as the most efficient ESS norms, as demonstrated in Ref.~\cite{ohtsuki2009indirect}.

On the other hand, when $q$ is sufficiently high but $\xi$ is low, the fraction of good players $h^{\ast}$ and the cooperation level remain high because defectors are misperceived as cooperators.
As a result, the equilibrium payoffs under CP norms are as high as those under CD norms.
The parameter region where CP norms are the most efficient is much larger than in the baseline model.
Even though defectors are not accurately identified, cooperation can be stably maintained with a high $\beta$.
The risk of strong punishment, despite its low probability, discourages players from defecting.

\section{Discussion}

In this study, we explored the role of costly punishment in the context of indirect reciprocity, with a particular focus on situations where the detection of defection is error-prone.
Our findings challenge the conclusions of previous studies, which suggested that costly punishment offers only a narrow margin of efficiency in promoting cooperation~\cite{ohtsuki2009indirect}.
Specifically, earlier research demonstrated that while costly punishment might contribute to evolutionary stability in noisy environments, it does so at the expense of significantly reducing overall payoffs.
Contrary to these earlier conclusions, our study shows that costly punishment can indeed be beneficial under certain conditions, particularly when other types of errors are present.

We focused on scenarios where defection might be misperceived as cooperation--—a situation that previous models have not adequately addressed.
These scenarios are realistic since defectors have a natural incentive to hide their defection and deceive others into perceiving their actions as cooperative.
Our results indicate that under these circumstances, costly punishment can facilitate evolutionarily stable cooperation without severely diminishing the overall level of cooperation.
This finding suggests that costly punishment plays a more effective role in indirect reciprocity than previously thought, especially in environments where accurately identifying defection is challenging.

From a mathematical perspective, the key to the effectiveness of costly punishment in our model lies in the asymmetric nature of the errors involved.
In traditional models of indirect reciprocity, errors are typically induced symmetrically; that is, the probability of a good action being misassessed as bad is equal to the probability of a bad action being misassessed as good.
However, in our model, these errors are induced asymmetrically, with the likelihood of misperceiving defection as cooperation differing from the likelihood of misperceiving cooperation as defection.
This asymmetry is critical because it allows populations to maintain a high level of cooperation even in the presence of noise, making CP norms significantly more efficient compared to the symmetric error model.
Even though defection is harder to detect under the asymmetric errors, the population can still effectively suppress free-riders via strong punishment, thereby sustaining cooperation.
We may interpret such an asymmetry in errors as a custom that the society intentionally implemented.
Sometimes the identification of actions may be imperfect due to noise or cost.
It is beneficial to assess uncertain players as good rather than randomly assigning them a good or bad label.
By doing so, the population can maintain a high level of cooperation.
To compensate for the risk of misidentifying defectors as cooperators, the society needs to implement a strong punishment system to deter potential defectors from exploitation.
This is how costly punishment can be effective in promoting cooperation in indirect reciprocity even if it incurs some cost.
Future studies on models with uncertain states like those in Ref.~\cite{nakamura2011indirect} may provide further insights.

While our findings provide new insights into the role of costly punishment in indirect reciprocity, several open questions remain for future research.
In this study, we focused on second-order social norms, which represent the simplest class of norms that can sustain evolutionarily stable cooperation.
Future studies should explore more complex strategies~\cite{santos2018social}, such as third-order social norms~\cite{ohtsuki2004should}, stochastic norms~\cite{murase2023indirect,schmid2021unified,schmid2021evolution}, and norms involving non-binary reputations~\cite{murase2022social,schmid2023quantitative,lee2021local}.
For example, in the realm of third-order social norms, the ``secondary sixteen'' norms have been shown to promote cooperation~\cite{murase2023indirect}, and it would be intriguing to investigate the effectiveness of costly punishment within these norms.

Another promising avenue for future research is the study of dual-reputation updates~\cite{murase2023indirect}, where not only the donor's reputation but also the recipient's reputation is updated.
Previous research suggests that dual-reputation updates make norms more error-tolerant and thus more efficient, albeit at the cost of requiring a higher benefit-to-cost ratio to maintain cooperation.
It would be interesting to examine whether costly punishment could reduce this required ratio while preserving the efficiency of dual-reputation systems.

Additionally, the private assessment models~\cite{hilbe2018indirect}, in which players' assessments of others may not always be synchronized, presents another interesting direction for future studies.
In such models, costly punishment may be particularly effective because it could help distinguish free-riders from those who engage in justified punishment.
A major open question in this area is how costly punishment and social norms can co-evolve initially~\cite{murase2024computational}.
We hypothesize that it may evolve in situations where observing actions is costly and error-prone, and punishment emerges as a way to suppress free-riding in this challenging environment.

In conclusion, our study demonstrates that costly punishment can be a valuable mechanism in sustaining cooperation in indirect reciprocity, particularly in scenarios where errors in detecting defection are present.
These findings open up new avenues for exploring more complex social norms and assessment models, potentially leading to a deeper understanding of the mechanisms that underpin cooperation in human societies.

\section*{Acknowledgments}
Y.M. acknowledges support from Japan Society for the Promotion of Science (JSPS) (JSPS KAKENHI; Grant no. 21K03362 and Grant no. 23K22087).
Y.M. is grateful to Akari Kakinuma for her assistance with numerical confirmation and for her inspiring discussions.

\bibliographystyle{unsrtnat}
\bibliography{indirect}

\begin{thebibliography}{32}
\providecommand{\natexlab}[1]{#1}
\providecommand{\url}[1]{\texttt{#1}}
\expandafter\ifx\csname urlstyle\endcsname\relax
  \providecommand{\doi}[1]{doi: #1}\else
  \providecommand{\doi}{doi: \begingroup \urlstyle{rm}\Url}\fi

\bibitem[Rand and Nowak(2012)]{rand:TCS:2013}
D.~G. Rand and M.~A. Nowak.
\newblock Human cooperation.
\newblock \emph{Trends in Cogn. Sciences}, 117:\penalty0 413--425, 2012.

\bibitem[Melis and Semmann(2010)]{melis:ptrs:2010}
A.~P. Melis and D.~Semmann.
\newblock How is human cooperation different?
\newblock \emph{Philosophical Transactions of the Royal Society B},
  365:\penalty0 2663--2674, 2010.

\bibitem[Nowak and Sigmund(2005)]{nowak2005evolution}
Martin~A Nowak and Karl Sigmund.
\newblock Evolution of indirect reciprocity.
\newblock \emph{Nature}, 437\penalty0 (7063):\penalty0 1291--1298, 2005.

\bibitem[Wedekind and Braithwaite(2002)]{wedekind2002long}
Claus Wedekind and Victoria~A Braithwaite.
\newblock The long-term benefits of human generosity in indirect reciprocity.
\newblock \emph{Current biology}, 12\penalty0 (12):\penalty0 1012--1015, 2002.

\bibitem[Sigmund(2012)]{sigmund2012moral}
Karl Sigmund.
\newblock Moral assessment in indirect reciprocity.
\newblock \emph{Journal of theoretical biology}, 299:\penalty0 25--30, 2012.

\bibitem[Santos et~al.(2021)Santos, Pacheco, and Santos]{santos2021complexity}
Fernando~P Santos, Jorge~M Pacheco, and Francisco~C Santos.
\newblock The complexity of human cooperation under indirect reciprocity.
\newblock \emph{Philosophical Transactions of the Royal Society B},
  376\penalty0 (1838):\penalty0 20200291, 2021.

\bibitem[Okada(2020)]{okada2020review}
Isamu Okada.
\newblock A review of theoretical studies on indirect reciprocity.
\newblock \emph{Games}, 11\penalty0 (3):\penalty0 27, 2020.

\bibitem[Fehr and G{\"a}chter(2002)]{fehr2002altruistic}
Ernst Fehr and Simon G{\"a}chter.
\newblock Altruistic punishment in humans.
\newblock \emph{Nature}, 415\penalty0 (6868):\penalty0 137--140, 2002.

\bibitem[De~Quervain et~al.(2004)De~Quervain, Fischbacher, Treyer,
  Schellhammer, Schnyder, Buck, and Fehr]{de2004neural}
Dominique J-F De~Quervain, Urs Fischbacher, Valerie Treyer, Melanie
  Schellhammer, Ulrich Schnyder, Alfred Buck, and Ernst Fehr.
\newblock The neural basis of altruistic punishment.
\newblock \emph{Science}, 305\penalty0 (5688):\penalty0 1254--1258, 2004.

\bibitem[Mathew and Boyd(2011)]{mathew2011punishment}
Sarah Mathew and Robert Boyd.
\newblock Punishment sustains large-scale cooperation in prestate warfare.
\newblock \emph{Proceedings of the National Academy of Sciences}, 108\penalty0
  (28):\penalty0 11375--11380, 2011.

\bibitem[Wu et~al.(2022)Wu, Luan, and Raihani]{wu2022reward}
Junhui Wu, Shenghua Luan, and Nichola Raihani.
\newblock Reward, punishment, and prosocial behavior: Recent developments and
  implications.
\newblock \emph{Current Opinion in Psychology}, 44:\penalty0 117--123, 2022.

\bibitem[Balafoutas et~al.(2014)Balafoutas, Nikiforakis, and
  Rockenbach]{balafoutas2014direct}
Loukas Balafoutas, Nikos Nikiforakis, and Bettina Rockenbach.
\newblock Direct and indirect punishment among strangers in the field.
\newblock \emph{Proceedings of the National Academy of Sciences}, 111\penalty0
  (45):\penalty0 15924--15927, 2014.

\bibitem[Ule et~al.(2009)Ule, Schram, Riedl, and Cason]{ule2009indirect}
Alja{\v{z}} Ule, Arthur Schram, Arno Riedl, and Timothy~N Cason.
\newblock Indirect punishment and generosity toward strangers.
\newblock \emph{Science}, 326\penalty0 (5960):\penalty0 1701--1704, 2009.

\bibitem[Guala(2012)]{guala2012reciprocity}
Francesco Guala.
\newblock Reciprocity: Weak or strong? what punishment experiments do (and do
  not) demonstrate.
\newblock \emph{Behavioral and brain sciences}, 35\penalty0 (1):\penalty0
  1--15, 2012.

\bibitem[Raihani and Bshary(2015)]{raihani2015reputation}
Nichola~J Raihani and Redouan Bshary.
\newblock The reputation of punishers.
\newblock \emph{Trends in ecology \& evolution}, 30\penalty0 (2):\penalty0
  98--103, 2015.

\bibitem[Raihani and Bshary(2019)]{raihani2019punishment}
Nichola~J Raihani and Redouan Bshary.
\newblock Punishment: one tool, many uses.
\newblock \emph{Evolutionary Human Sciences}, 1:\penalty0 e12, 2019.

\bibitem[Li et~al.(2018)Li, Jusup, Wang, Li, Shi, Podobnik, Stanley, Havlin,
  and Boccaletti]{li2018punishment}
Xuelong Li, Marko Jusup, Zhen Wang, Huijia Li, Lei Shi, Boris Podobnik,
  H~Eugene Stanley, Shlomo Havlin, and Stefano Boccaletti.
\newblock Punishment diminishes the benefits of network reciprocity in social
  dilemma experiments.
\newblock \emph{Proceedings of the National Academy of Sciences}, 115\penalty0
  (1):\penalty0 30--35, 2018.

\bibitem[Hauser et~al.(2014)Hauser, Nowak, and Rand]{hauser2014punishment}
Oliver~P Hauser, Martin~A Nowak, and David~G Rand.
\newblock Punishment does not promote cooperation under exploration dynamics
  when anti-social punishment is possible.
\newblock \emph{Journal of theoretical biology}, 360:\penalty0 163--171, 2014.

\bibitem[Rand and Nowak(2011)]{rand2011evolution}
David~G Rand and Martin~A Nowak.
\newblock The evolution of antisocial punishment in optional public goods
  games.
\newblock \emph{Nature communications}, 2\penalty0 (1):\penalty0 434, 2011.

\bibitem[Ohtsuki et~al.(2009)Ohtsuki, Iwasa, and Nowak]{ohtsuki2009indirect}
Hisashi Ohtsuki, Yoh Iwasa, and Martin~A Nowak.
\newblock Indirect reciprocity provides only a narrow margin of efficiency for
  costly punishment.
\newblock \emph{Nature}, 457\penalty0 (7225):\penalty0 79, 2009.

\bibitem[Nakamura and Masuda(2011)]{nakamura2011indirect}
Mitsuhiro Nakamura and Naoki Masuda.
\newblock Indirect reciprocity under incomplete observation.
\newblock \emph{PLoS Comput Biol.}, 7\penalty0 (7):\penalty0 e1002113, 2011.

\bibitem[Ohtsuki and Iwasa(2004)]{ohtsuki2004should}
Hisashi Ohtsuki and Yoh Iwasa.
\newblock How should we define goodness? -- reputation dynamics in indirect
  reciprocity.
\newblock \emph{J. Theor. Biol.}, 231\penalty0 (1):\penalty0 107--120, 2004.

\bibitem[Ohtsuki and Iwasa(2006)]{ohtsuki2006leading}
Hisashi Ohtsuki and Yoh Iwasa.
\newblock The leading eight: social norms that can maintain cooperation by
  indirect reciprocity.
\newblock \emph{J. Theor. Biol.}, 239\penalty0 (4):\penalty0 435--444, 2006.

\bibitem[Murase and Hilbe(2023)]{murase2023indirect}
Yohsuke Murase and Christian Hilbe.
\newblock Indirect reciprocity with stochastic and dual reputation updates.
\newblock \emph{PLOS Computational Biology}, 19\penalty0 (7):\penalty0
  e1011271, 2023.

\bibitem[Santos et~al.(2018)Santos, Santos, and Pacheco]{santos2018social}
Fernando~P Santos, Francisco~C Santos, and Jorge~M Pacheco.
\newblock Social norm complexity and past reputations in the evolution of
  cooperation.
\newblock \emph{Nature}, 555\penalty0 (7695):\penalty0 242--245, 2018.

\bibitem[Schmid et~al.(2021{\natexlab{a}})Schmid, Chatterjee, Hilbe, and
  Nowak]{schmid2021unified}
Laura Schmid, Krishnendu Chatterjee, Christian Hilbe, and Martin~A Nowak.
\newblock A unified framework of direct and indirect reciprocity.
\newblock \emph{Nat. Hum. Behav.}, 5:\penalty0 1292, 2021{\natexlab{a}}.

\bibitem[Schmid et~al.(2021{\natexlab{b}})Schmid, Shati, Hilbe, and
  Chatterjee]{schmid2021evolution}
Laura Schmid, Pouya Shati, Christian Hilbe, and Krishnendu Chatterjee.
\newblock The evolution of indirect reciprocity under action and assessment
  generosity.
\newblock \emph{Scientific Reports}, 11\penalty0 (1):\penalty0 17443,
  2021{\natexlab{b}}.

\bibitem[Murase et~al.(2022)Murase, Kim, and Baek]{murase2022social}
Yohsuke Murase, Minjae Kim, and Seung~Ki Baek.
\newblock Social norms in indirect reciprocity with ternary reputations.
\newblock \emph{Scientific Reports}, 12\penalty0 (1):\penalty0 455, 2022.

\bibitem[Schmid et~al.(2023)Schmid, Ekbatani, Hilbe, and
  Chatterjee]{schmid2023quantitative}
Laura Schmid, Farbod Ekbatani, Christian Hilbe, and Krishnendu Chatterjee.
\newblock Quantitative assessment can stabilize indirect reciprocity under
  imperfect information.
\newblock \emph{Nature Communications}, 14\penalty0 (1):\penalty0 2086, 2023.

\bibitem[Lee et~al.(2021)Lee, Murase, and Baek]{lee2021local}
Sanghun Lee, Yohsuke Murase, and Seung~Ki Baek.
\newblock Local stability of cooperation in a continuous model of indirect
  reciprocity.
\newblock \emph{Scientific Reports}, 11\penalty0 (1):\penalty0 14225, 2021.

\bibitem[Hilbe et~al.(2018)Hilbe, Schmid, Tkadlec, Chatterjee, and
  Nowak]{hilbe2018indirect}
Christian Hilbe, Laura Schmid, Josef Tkadlec, Krishnendu Chatterjee, and
  Martin~A Nowak.
\newblock Indirect reciprocity with private, noisy, and incomplete information.
\newblock \emph{Proc. Natl. Acad. Sci. USA}, 115\penalty0 (48):\penalty0
  12241--12246, 2018.

\bibitem[Murase and Hilbe(2024)]{murase2024computational}
Yohsuke Murase and Christian Hilbe.
\newblock Computational evolution of social norms in well-mixed and
  group-structured populations.
\newblock \emph{Proceedings of the National Academy of Sciences}, 121\penalty0
  (33):\penalty0 e2406885121, 2024.

\end{thebibliography}

\end{document}